\documentclass{nature}
\usepackage{graphicx}
\makeatletter
\let\saved@includegraphics\includegraphics
\AtBeginDocument{\let\includegraphics\saved@includegraphics}
\renewenvironment*{figure}{\@float{figure}}{\end@float}
\makeatother

\usepackage[english]{babel}
\usepackage[utf8x]{inputenc}
\usepackage[T1]{fontenc}

\usepackage[a4paper,top=3cm,bottom=2cm,left=3cm,right=3cm,marginparwidth=1.75cm]{geometry}

\usepackage{url}
\usepackage[section]{placeins}
\usepackage{booktabs}
\usepackage{amsmath}
\usepackage{subcaption}
\usepackage[colorinlistoftodos]{todonotes}
\usepackage[colorlinks=true, allcolors=blue]{hyperref}
\usepackage{lineno}
\title{Forecasting Internally Displaced Population Migration Patterns in Syria and Yemen}
\author{Benjamin Q. Huynh\thanks{Address correspondence to \href{mailto:benhuynh@stanford.edu}{benhuynh@stanford.edu}.} \ and Sanjay Basu}

\begin{document}
\maketitle

\begin{abstract}
Armed conflict has led to an unprecedented number of internally displaced persons (IDPs) - individuals who are forced out of their homes but remain within their country. IDPs often urgently require shelter, food, and healthcare, yet prediction of when large fluxes of IDPs will cross into an area remains a major challenge for aid delivery organizations.\cite{feller2006unhcr,UNHCRreport} Accurate forecasting of IDP migration would empower humanitarian aid groups to more effectively allocate resources during conflicts. We show that monthly flow of IDPs from province to province in both Syria and Yemen can be accurately forecasted one month in advance, using publicly available data. We model monthly IDP flow using data on food price, fuel price, wage, geospatial, and news data. We find that machine learning approaches can more accurately forecast migration trends than baseline persistence models. Our findings thus potentially enable proactive aid allocation for IDPs in anticipation of forecasted arrivals.
%
%
%
%
\end{abstract}

\section*{Main}

Armed conflict has contributed to an alarming rate of migration, with over 65 million forcibly displaced people worldwide.\cite{UNHCRreport} Of those displaced, about 40 million are considered internally displaced persons (IDPs) - individuals who are forced out of their homes but remain within their country. IDPs often require aid in the form of food, shelter, or health care. However, because much of internal displacement arises from regional instability for which local authorities are either unwilling or unequipped to provide aid, it is rare for IDPs to find support from their governments.\cite{goodwin2007refugee} 

Humanitarian response to IDPs is instead typically provided by international nongovernmental organizations, including the United Nations and its affiliate agencies.\cite{feller2006unhcr} Such aid groups face many logistical challenges in providing support to IDPs in conflict-rife zones, one of which is resource allocation across many possible sites to which IDPs may migrate. Given the chaos and unpredictability of conflict zones, it is difficult to anticipate when and where IDPs will arrive, so it is unclear which shelters and camps will reach capacity soonest, and where supplies and workers should be sent. At present, allocations are ad hoc and often delayed. It would therefore be valuable to anticipate IDP flow, so that aid groups can proactively prepare to distribute resources in an anticipatory manner. 

Previous attempts to forecast forced migration have been limited by both data and methods. Prior literature has focused largely on inferring risk factors for forced migration rather than on accurate forecasting for operational use.\cite{riskfactor1,riskfactor2,riskfactor3,riskfactor4} The few studies focused on forecasting involve heuristic-based simulations of theoretical migration patterns with strong assumptions\cite{heur1,heur3,heur4} or linear statistical models with very few available predictors,\cite{stat1,stat2,stat3} resulting in poor predictive performance or unreliable predictions on external data. Furthermore, no study to our knowledge has modeled IDP flow, which is more granular and noisy than international migration despite being more common. Historically, IDP flow data have been delayed, missing, and sometimes anecdotal rather than consistently collected, due to difficulties in obtaining reliable data. 

In recent years, internal displacement task forces have been established to collect on-site data on IDP flow through surveys, registrations, and site monitoring, all of which are triangulated and verified through multiple sources.\cite{idmc} Additionally, other public data sources have emerged, providing potential predictors of flow, including market prices for staple goods, wages, and conflict events.\cite{acled,icews,wfp} These data may permit detailed forecasting of IDP flow for the first time.
%
%
%
%
%
%

Here, we develop a model of monthly flow of IDPs, using data on risk factors and migration from province to province in both Syria and Yemen. Since conflict began in 2011, both countries have had populations experiencing severe levels of displacement; there are over 7 million IDPs in Syria and 2.5 million in Yemen.\cite{idmc} By integrating the aforementioned data and deriving predictive features from them, we are able to construct  statistical models and evaluate their predictive performance for forecasting IDP flow one month in advance. 
%
%
%


\begin{figure}
\includegraphics[width = \textwidth]{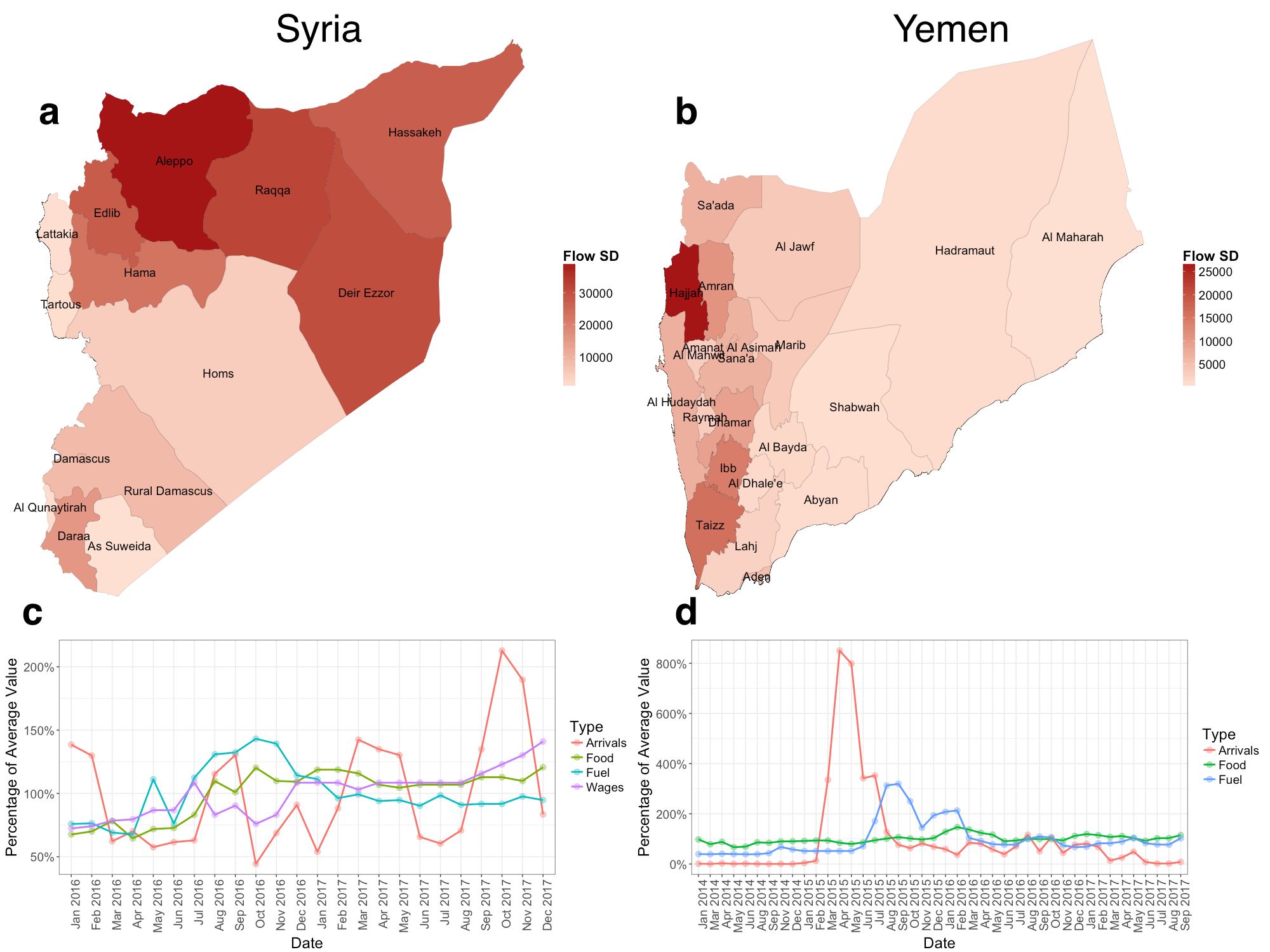}
\caption{Measurement variability over time for Syria and Yemen. a,b: Provinces of each country color coded by standard deviation of IDP arrivals aggregated over time. Darker shades indicate larger variability in IDP flow for a given province. c,d: Country level statistics on IDP arrivals, food prices, fuel prices, and wages over time for Syria and Yemen. Values are presented as percentages of their historical averages. Wage data are unavailable for Yemen.}
\label{fig:map}
\end{figure}

The data from both Syria and Yemen revealed large province-to-province and month-to-month variations in IDP flow, as well as in key covariates we studied for prediction: food prices, fuel prices, and wages. The relative standard deviations of IDP flow were extremely large for both Syria and Yemen (389\% and 517\% respectively), suggesting high variability in flow across provinces and months (Figure \ref{fig:map}, Extended Data Table \ref{tab:descriptive}). The price data also yielded large relative standard deviations: 34\%, 52\%, and 27\%, for food prices, fuel prices, and wages in Syria; 55\% and 142\% for food and fuel prices in Yemen.

%
%
%
%
%
%
%
%

\begin{figure}
\includegraphics[width = \textwidth]{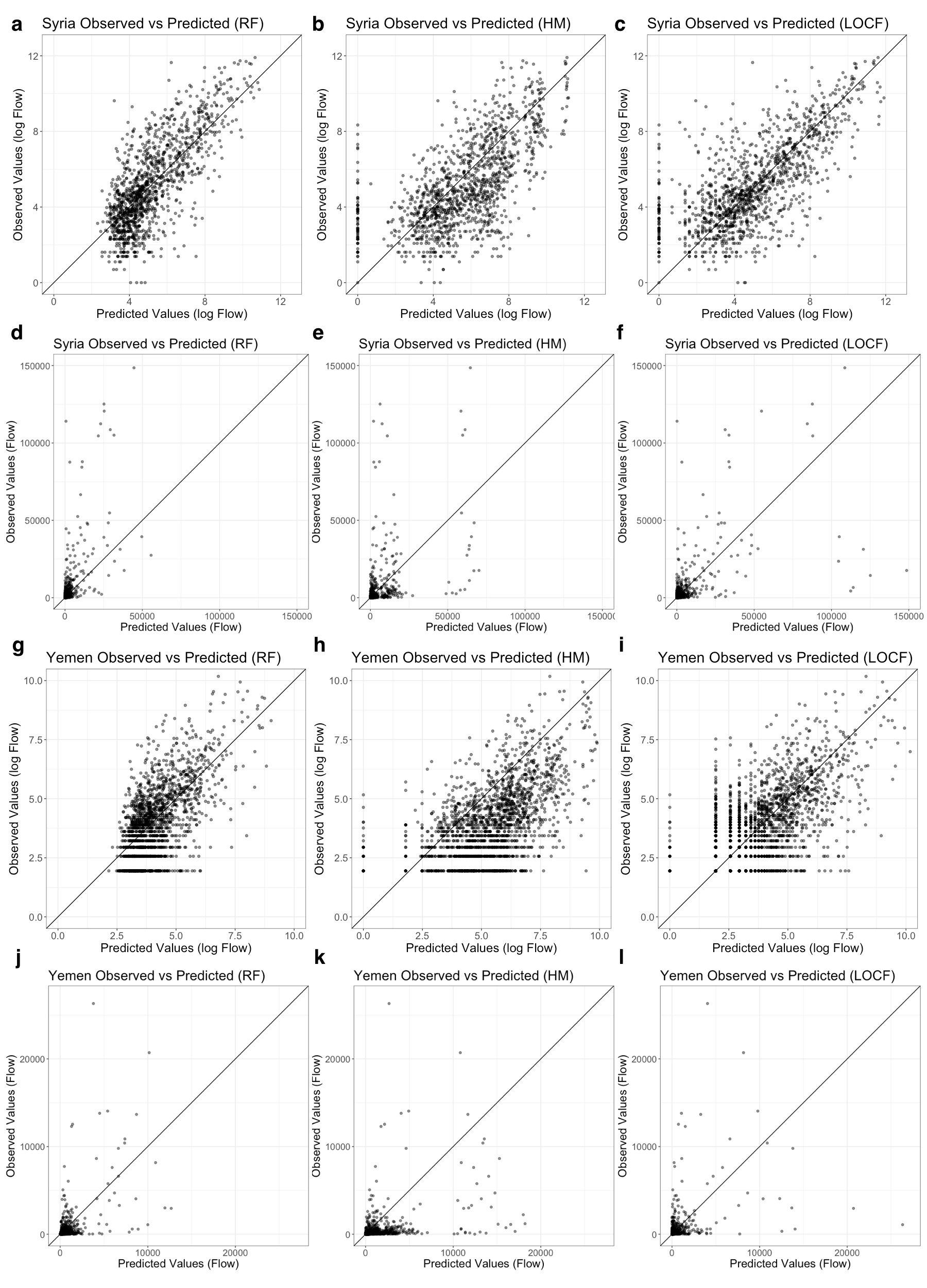}
\caption{Observed vs predicted values for IDP flow (d-f, j-l) and log flow (a-c, g-i) aggregated across all available months and provinces. Left column (a,d,g,j) depicts plots from a random forest model (RF), middle column (b,e,h,k) depicts historical mean values (HM), and right column (c,f,i,l) depicts last observations carried forward (LOCF).}
%
%
%
%
%
\label{fig:obsvspred}
\end{figure}

The baseline persistence models we tested, historical mean and last observation carried forward (LOCF), were able to capture trends of IDP flow and log-flow within each province with root mean square error (RMSE = 10587 and 10661 for Syria HM and LOCF, 1332 and 1413 for Yemen HM and LOCF ) and mean absolute error (MAE = 3066 and 2577 for Syria HM and LOCF, 288 and 326 for Yemen HM and LOCF) values that are moderately low, but with poor $R^{2}$ ($R^{2}$ = 0.24 and 0.34 for Syria HM and LOCF, 0.08 and 0.17 for Yemen HM and LOCF) values (Table \ref{tab:results}).  Because the persistence models relied solely on historical data, they were unable to provide forecasts for regions for which there had previously been no IDP arrivals. As shown in Figure \ref{fig:obsvspred}, the historical mean and LOCF models produced many erroneous zero predictions. Additionally, both persistence models performed poorly in terms of sign accuracy (63\% and 59\% for Syria HM and LOCF, 67\% and 60\% for Yemen HM and LOCF), a measure of how well the models forecast whether a given observation would be an increase or decrease in flow compared to the previous month.
%
%
%
%

By comparison, the linear mixed-effects model and particularly the machine learning algorithms we trained outperformed historical mean and LOCF forecasts in terms of RMSE, MAE, and $R^2$ for both predicting flow and log-flow, as well as sign accuracy (Table \ref{tab:results}, Extended Data Figure \ref{fig:sankeyLog}). Furthermore, the mixed-effects and machine learning models were able to make predictions for regions without previous data, avoiding erroneous zero predictions (Figure \ref{fig:obsvspred}). The mixed-effects model and machine learning algorithms had similar predictive performances, though the random forest machine learning algorithm appeared slightly better overall than the others across both countries. The random forest specifically outperformed LOCF in terms of RMSE of log flow by 26\% and 17\% for the Syria and Yemen datasets, respectively.

We observed that the random forest, our best model, captured overall trends of IDP arrivals for each province but occasionally failed to capture sudden spikes in displacement (Figure \ref{fig:destflow}, Figure \ref{fig:obsvspred}). Our model obtained a 70\% and 74\% sign accuracy for Syria and Yemen, respectively (Table \ref{tab:results}).  These are relatively high values ($\pm$ 2\% sign accuracy compared to other machine learning models we tested), but they also suggest room for improvement in absolute terms of detecting spikes.  This is likely because our features are unable to fully characterize when spikes occur. For example, our conflict intensity metric is determined by how many armed conflict events occur in a month, but does not consider the magnitude of the armed conflict event. It should be noted that in comparison, baseline persistence methods fundamentally perform poorly at detecting large spikes in displacement since they simply project past data.


%
%
%
%
%
%

\begin{figure}
\includegraphics[width=\textwidth]{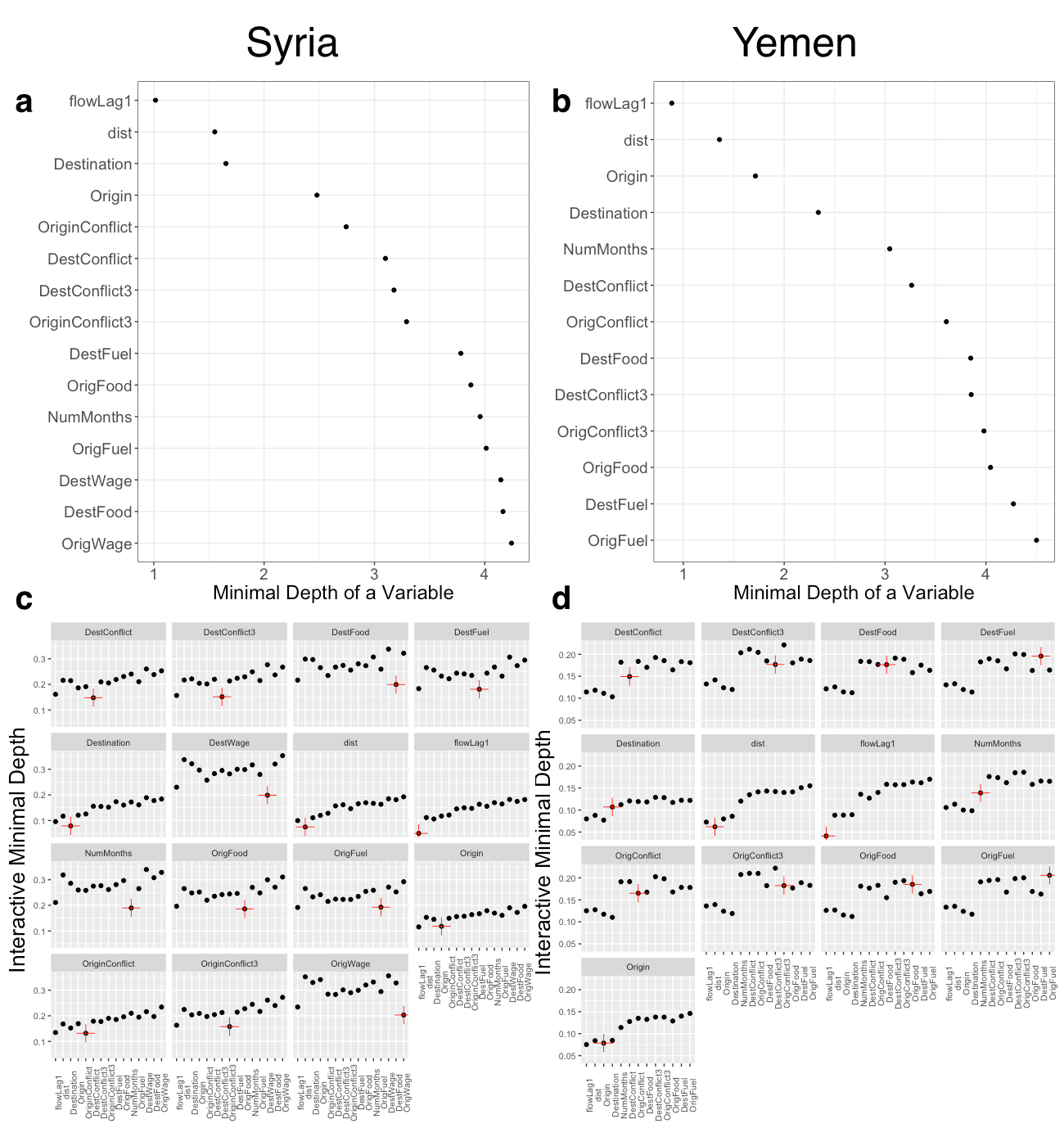}
\caption{a,b: Random forest minimal depth variables in ranked order for Syria (a) and Yemen (b), with the most important variables at the top. Smaller values of minimal depth indicate a stronger impact on the forest prediction. c,d: Minimal depth variable interactions for Syria (c) and Yemen (d). Red cross indicates the reference variable for each panel. Higher levels of interactivity are indicated by lower levels of minimal depth.}
\label{fig:minDepth}
\end{figure}

Interpretation of the random forest model yielded sensible results, suggesting our models are finding patterns within the data and not just fitting to noise. The minimal depth levels, a measurement of how much impact a given variable had on the final prediction, appeared plausible for both datasets (Extended Data Figure \ref{fig:minDepth}). The autoregressive term is unsurprisingly the strongest predictor - we expected last month's flow to be a good estimate of this month's flow. Distance was the second strongest predictor - most IDPs become displaced within their home province (Extended Data Figure \ref{fig:sankeyLog}), so we expect shorter distances between the origin and destination provinces to be associated with larger flow. Food prices and conflict intensity are also strong predictors, likely due to famine and severe civil conflict in both countries.
%
%


Our main result, that IDP flow from province to province can be forecast using mixed-effects or machine learning techniques, revealed that a data-driven approach to modeling IDP migration patterns during a crisis was viable, despite the substantial heterogeneity and variability in flow over time and across provinces. 


The limitations of this work are largely related to the quality of the available data. There is substantial uncertainty inherent to the datasets we used: the ground truth for IDP flow numbers, conflict events, prices, and wages are all subject to the unreliability of on-site data collection. The IDP flow values also lack potentially valuable disaggregated information, such as age or sex, or more granular information, such as daily flow (instead of monthly) or subdistrict-level flow (instead of provinces). There were also substantial amounts of missing data from the price dataset for Yemen (see Methods) that were imputed, possibly resulting in bias.
%
%
%
%
%
%
%

Future work could involve incorporating new kinds of data into our models. This might entail working closely with humanitarian organizations to directly integrate our models into the task of resource allocation. Other approaches could involve obtaining new datasets, such as acquiring annotated satellite imagery, cell phone data, or relevant social media posts and adding them to our models. 

%
%
%

The plight of global forced migration is projected to only worsen in coming years, both in terms of current crises worsening and new crises arising.\cite{haas} We believe findings such as ours can help establish a system of informed data-driven responses to such urgent situations.
%
%
%

\begin{figure}[ht]
\includegraphics[width = \textwidth]{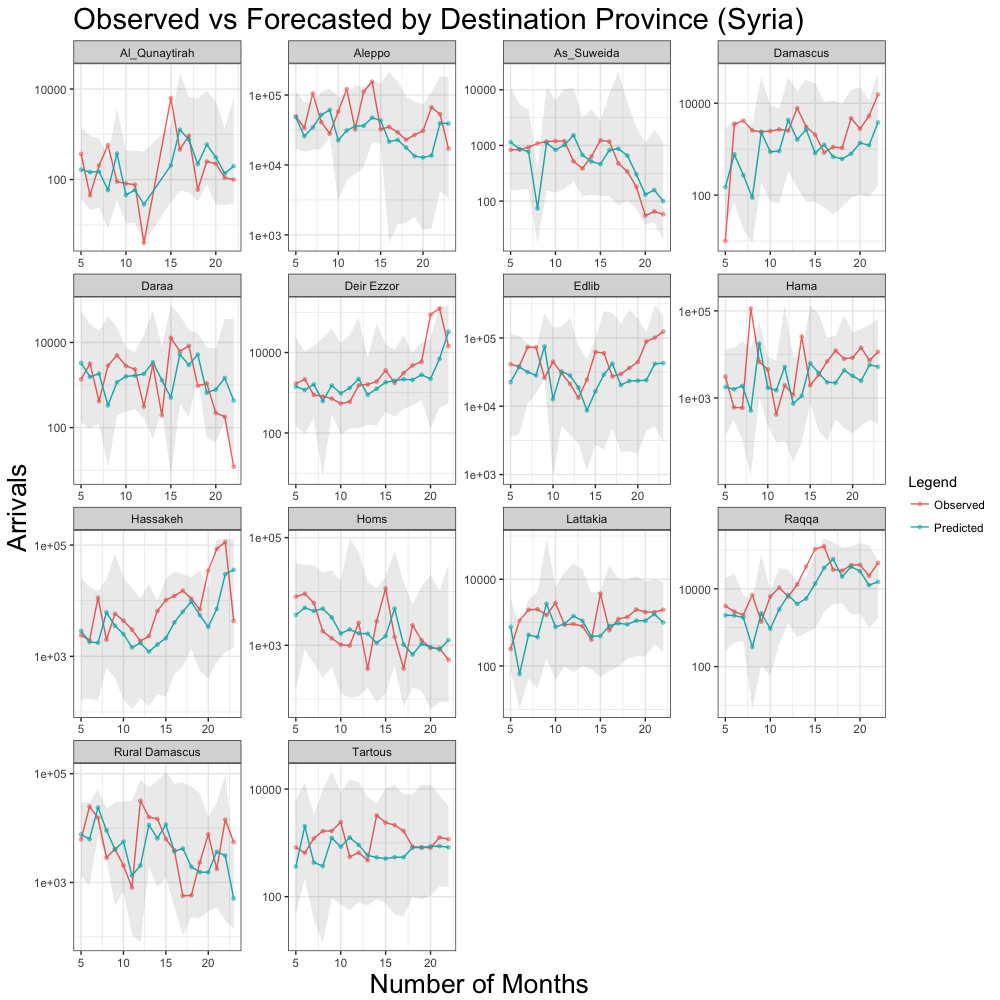}
\caption{Observed and forecasted number of IDP arrivals in each province by month for Syria and Yemen. Forecasts for each month were made from a random forest model trained on data from prior months. Gray shaded regions denote 95\% prediction intervals determined by the quantiles of the individual trees for each prediction.}
\label{fig:destflow}
\end{figure}
\begin{figure}[ht]\ContinuedFloat
\includegraphics[width = \textwidth]{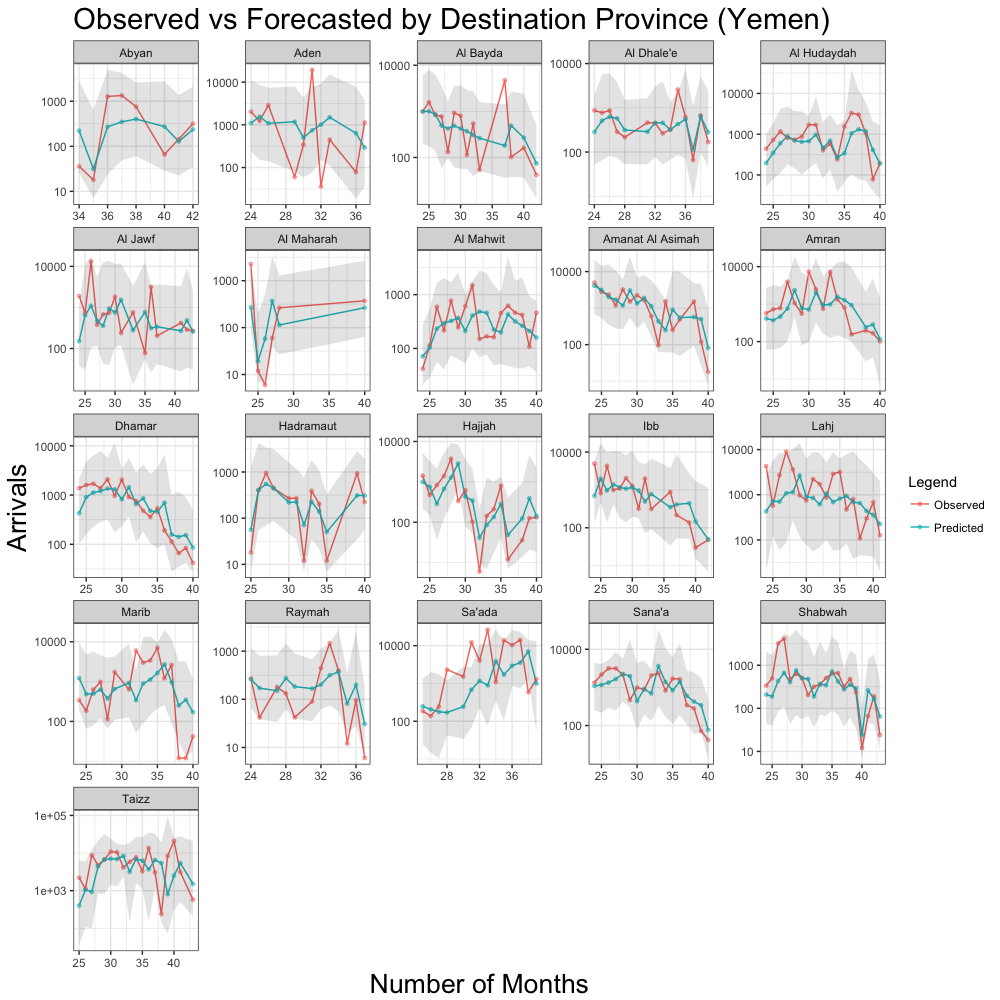}
\caption{Observed and forecasted number of IDP arrivals in each province by month for Syria and Yemen. Forecasts for each month were made from a random forest model trained on data from prior months. Gray shaded regions denote 95\% prediction intervals determined by the quantiles of the individual trees for each prediction. (cont.)}
\end{figure}

\begin{table}
\begingroup
\small
\caption{Predictive performance of forecasting methods for Syria (a) and Yemen (b) on both flow and log-flow. HM denotes historical mean, LOCF denotes last observation carried forward, LMM denotes linear mixed effects model, SVM denotes support vector machine,  RF denotes random forest, MERF denotes mixed-effects random forest, XGB denotes gradient boosting, and MLP denotes multi-layer perceptron. RMSE is root mean squared error, MAE is mean absolute error, and $R^2$ is the coefficient of determination.}
\begin{subtable}{\linewidth}
\centering
\caption{Syria predictive performance.}
\begin{tabular}{lrrrrrrr}
  \hline
Model & RMSE & MAE & $R^{2}$ & RMSE (log) & MAE (log) & $R^{2}$ (log) & Sign Acc. \\ 
  \hline
HM & 10587.07 & 3066.02 & 0.24 & 2.15 & 1.66 & 0.38 & 0.63 \\ 
LOCF & 10660.7 & 2577.37 & 0.34 & 2.01 & 1.44 & 0.46 & 0.59 \\ 
LMM & 10074.47 & 2370.81 & 0.31 & 1.55 & 1.19 & 0.56 & \textbf{0.70} \\ 
SVM & 10292.38 & 2383.21 & 0.26 & 1.53 & 1.16 & 0.57 & \textbf{0.70} \\ 
RF & \textbf{9576.61} & \textbf{2237.73} & \textbf{0.45} & \textbf{1.49} & \textbf{1.14} & \textbf{0.59} & \textbf{0.70} \\
MERF & 9627.89 & 2304.97 & 0.34 & 1.53 & 1.18 & 0.57 & \textbf{0.70} \\
XGB & 9760.46 & 2351.41 & 0.35 & 1.59 & 1.23 & 0.53 & 0.68 \\
MLP & 10283.04 & 2378.43 & 0.35 & 1.59 & 1.23 & 0.53 & 0.68 \\
   \hline
\end{tabular}
\end{subtable}
\begin{subtable}{\linewidth}
\centering
\caption{Yemen predictive performance.}
\begin{tabular}{lrrrrrrr}
  \hline
Model & RMSE & MAE & $R^{2}$ & RMSE (log) & MAE (log) & $R^{2}$ (log) & Sign Acc. \\ 
  \hline
HM & 1332.29 & 287.78 & 0.08 & 2.10 & 1.75 & 0.30 & 0.67 \\ 
LOCF & 1413.30 & 325.92 & 0.17 & 1.48 & 1.13 & 0.33 & 0.60 \\ 
LMM & 1175.50 & 276.59 & 0.17 & 1.31 & 1.02 & 0.37 & 0.73 \\ 
SVM & 1149.05 & 254.37 & \textbf{0.22} & 1.37 & 1.06 & 0.33 & 0.74 \\ 
RF & \textbf{1140.01} & \textbf{247.05} & 0.21 & \textbf{1.23} & \textbf{0.98} & \textbf{0.39} & 0.74 \\
MERF & 1161.15 & 250.41 & 0.19 & 1.25 & 0.98 & 0.38 & 0.75\\
XGB & 1236.94 & 258.51 & 0.12 & 1.27 & 0.98 & 0.37 & \textbf{0.76} \\
MLP & 1588.56 & 329.39 & 0.10 & 1.44 & 1.09 & 0.26 & 0.72 \\
   \hline
\end{tabular}
\end{subtable}
\label{tab:results}
\endgroup
\end{table}

\section*{Methods}

\subsection{Data}

We obtained monthly IDP flow data for Syria from a publicly available dataset provided by the United Nations Office for the Coordination of Humanitarian Affairs.\cite{OCHASyria} For Yemen, we obtained monthly IDP flow from a public dataset provided by the International Organization for Migration.\cite{yemen} The Syria dataset spans from January 2016 to December 2017; the Yemen dataset spans from January 2014 to September 2017. We determine each observation within our dataset to be each unique grouping of month, origin province, and destination province. The Syria and Yemen datasets thus contain 1505 and 3563 observations, respectively.
%
%

We obtained monthly food prices, fuel prices, and wages for both Syria and Yemen from the World Food Programme’s global food price database.\cite{wfp} The dataset contains values for commodities such as cheese, wheat, diesel, and gas. We calculated the median value for commodities within the categories of food, fuel, and wages. The values were originally recorded across distinct districts and marketplaces within governorates, so we calculated the medians of all values across governorates per month to get governorate-level values for the model. The global food price database is updated monthly and spans from the early 2000s to December 2017. We used within-month median imputation for missing values. 6\% of values were missing for Syria price data, and 46\% of values were missing for Yemen price data. Wage data were only available for July 2016 onward for Yemen, so we excluded it from analysis.
%
%
%
%
%
%
%
%

We obtained conflict data from two separate sources: The Integrated Conflict Early Warning System (ICEWS) dataset\cite{icews} and the Armed Conflict Location and Event Data (ACLED) collection\cite{acled}. The ICEWS dataset consists of political events across the globe and is publicly available for data spanning from 1995 to 2016. We take the subset of ICEWS events based on codes that correspond to armed conflict events. The ACLED dataset consists of global armed conflict event data; the Middle East ACLED data span from 2017 to May 2018. In order to featurize the conflict data for a statistical model, we created a “conflict intensity” feature. Conflict intensity was determined by the number of violent events per month for a given governorate, and then scaled to zero mean and unit variance. Scaling was done at the dataset level (separately for ICEWS and ACLED) to account for potential frequency biases in data collection between the two datasets.
%
%
%
%
%
%
%

We also created a distance metric by taking the coordinates of each province and calculating the Haversine distance\cite{haversine} between each province pair.
%
%
%

\subsection{Models}

For each model, our covariates consisted of monthly features derived from the aforementioned datasets. We used monthly data from both the origin and destination for each destination-origin pair to model the "push and pull" factors\cite{pushpull} of migration. Our model covariates for each observation were thus the date, monthly food prices, fuel prices, wages, and conflict intensity from both the origin and destination, as well as the distance between the origin and destination. To account for the fact that marketplace data would take time to be collected and shared in a real-time setting, we use food/fuel prices and wages with a three month lag instead of more recent data. For example, we use prices and wages from January 2017 to predict flow for April 2017. Conflict data are updated more frequently, so we use the previous month's conflict intensity metric for each observation. However, we also include a three month lagged conflict intensity variable to account for interactions with price data.  An autoregressive term, the previous month's IDP flow for a given origin-destination pair, is also included.
%
%
%
%

For our linear mixed effects model, we used a three-level structure with random slopes and intercepts.\cite{hedeker2006longitudinal} For notation, we defined $i = 1, ... , N$ origin provinces, $j = 1, ... , n_{i}$ origin-destination pairs, and $k = 1,..., n_{ij}$ monthly observations for each origin-destination pair.  Our formulation of the model was as follows: \[y_{i} = X_{i}\beta + Z_{i}v_{i} + \epsilon_{i}\] $X_{i}$ was the known design matrix for the fixed effects, $\beta$ was the unknown vector of regression coefficients, $Z_{i}$ was the known design matrix for the random effects, $v_{i}$ was the unknown vector of random effects with $v_{ij} \sim N(0,\Sigma_{v})$, and $\epsilon_{i}$ was the error term vector with $e_{ijk} \sim N(0,\sigma^{2})$.

We trained a support vector regression model\cite{svm},  a machine learning algorithm that seeks to find a function $f(x)$ that approximates $y$ by minimizing a loss function that ignores errors within a given distance $\epsilon$ of the true values.  We specified it with a polynomial kernel $K(x,y) = (x^{T}y + c)^{d}$. Hyperparameters $c$ and $d$ were selected through five-fold cross-validation on a training set.

We trained a random forest\cite{randomforest}, an algorithm that trains a large number of individual decision trees and takes the mean output as the prediction. We tuned the optimal number of variables randomly sampled at each split through five-fold cross-validation. We also trained a mixed-effects random forest\cite{merf}, with a similar specification to our linear mixed effects model: $y_{i} = f(X) + Z_{i}v_{i} + \epsilon_{i}$, where $f(X)$ was a standard random forest model.
%
%

We also used a tree boosting method, XGBoost\cite{xgboost}, which forms an ensemble of regression trees and builds a model in stages during training. The hyperparameters tuned through five-fold cross-validation were maximum tree depth, step size shrinkage, subsample ratio of columns (by tree), and subsample ratio of the training instance.
%
%

We trained a multi-layer perceptron (MLP), which is a class of feedforward deep neural networks.\cite{haykin2004comprehensive} Briefly, MLPs consists of layers of nodes, where each node is a neuron with a nonlinear activation function; the resulting network is thus a nonlinear function approximator. We specified our MLP with two hidden layers and rectifiers as activation functions. We selected the number of nodes through five-fold cross-validation.

For comparison, we also used baseline persistence methods of last observation carried forward (LOCF) and historical mean. Both methods are done from within origin-destination pairs. More specifically, historical mean was calculated as \[\hat{y}_{ijk} = (y_{ij1} + ... + y_{ijn_{ij}})/n_{ij}\] and LOCF was calculated as \[\hat{y}_{ijk} = y_{ijk-1}\]

We evaluated our models by forecasting out-of-sample one month ahead using a rolling origin. We started at month 5 for Syria (out of 24) and month 23 for Yemen (out of 44) - these are the time points at which all origin and destination provinces became present in the dataset. We use root mean squared error, mean absolute error, $R^{2}$, and sign accuracy as metrics for evaluation. Sign accuracy is a metric we introduce that measures how well a model can predict whether a given observation will be an increase or decrease in flow from the previous month. Specifically, we measure this as a binary classifier evaluation at each observation: an observation is a 1 if it is an increase in flow, and a 0 otherwise. We thus refer to the resulting binary classification accuracy as sign accuracy. We test our methods both on log-flow and flow (see Supplementary Information).

\subsection{Data Availability}
The IDP flow data for Yemen can be found at the International Organization for Migration's data repository (\url{https://displacement.iom.int/yemen}). The price data and Syria IDP flow data we used can be found at the Humanitarian Data Exchange's data repository (\url{https://data.humdata.org/}). The ACLED conflict data can be found at \url{https://www.acleddata.com/}. The ICEWS conflict data can be found at \url{https://dataverse.harvard.edu/dataset.xhtml?persistentId=doi:10.7910/DVN/28075}. We have also made all raw and processed datasets we used available at \url{https://github.com/benhuynh/migrationPatterns/tree/master/data}.

\subsection{Code Availability}
All code used for analysis and and figures are publicly available at \url{https://github.com/benhuynh/migrationPatterns}.


\section*{Author Information}
\subsection{Contributions}
B.Q.H. and S.B. designed the study and wrote the paper. B.Q.H. assembled and cleaned the dataset, and performed the analyses. Both authors discussed the results and commented on the manuscript.
\subsection{Competing Interests}
The authors declare no competing interests.
\subsection{Correspondence}
Correspondence to \href{mailto:benhuynh@stanford.edu}{B. Huynh}.

\section*{Supplementary Information}
\subsection{Data}

There exist two kinds of missing data in the IDP flow data: missing observations due to zero counts and missing observations due to some areas being inaccessible to surveyors. Despite the missing data, we opted for a complete case analysis instead of imputing the flow data. Our rationale is that if an area is inaccessible to surveyors, then it is also most likely inaccessible to humanitarian aid, so forecasting movements to those areas is not useful for assisting aid groups. Furthermore, it is unclear which missing observations represent zero flow and which ones represent unrecorded data, so imputation is not sensible. 
%
%
%
%

\begin{figure}
\includegraphics[width=\textwidth]{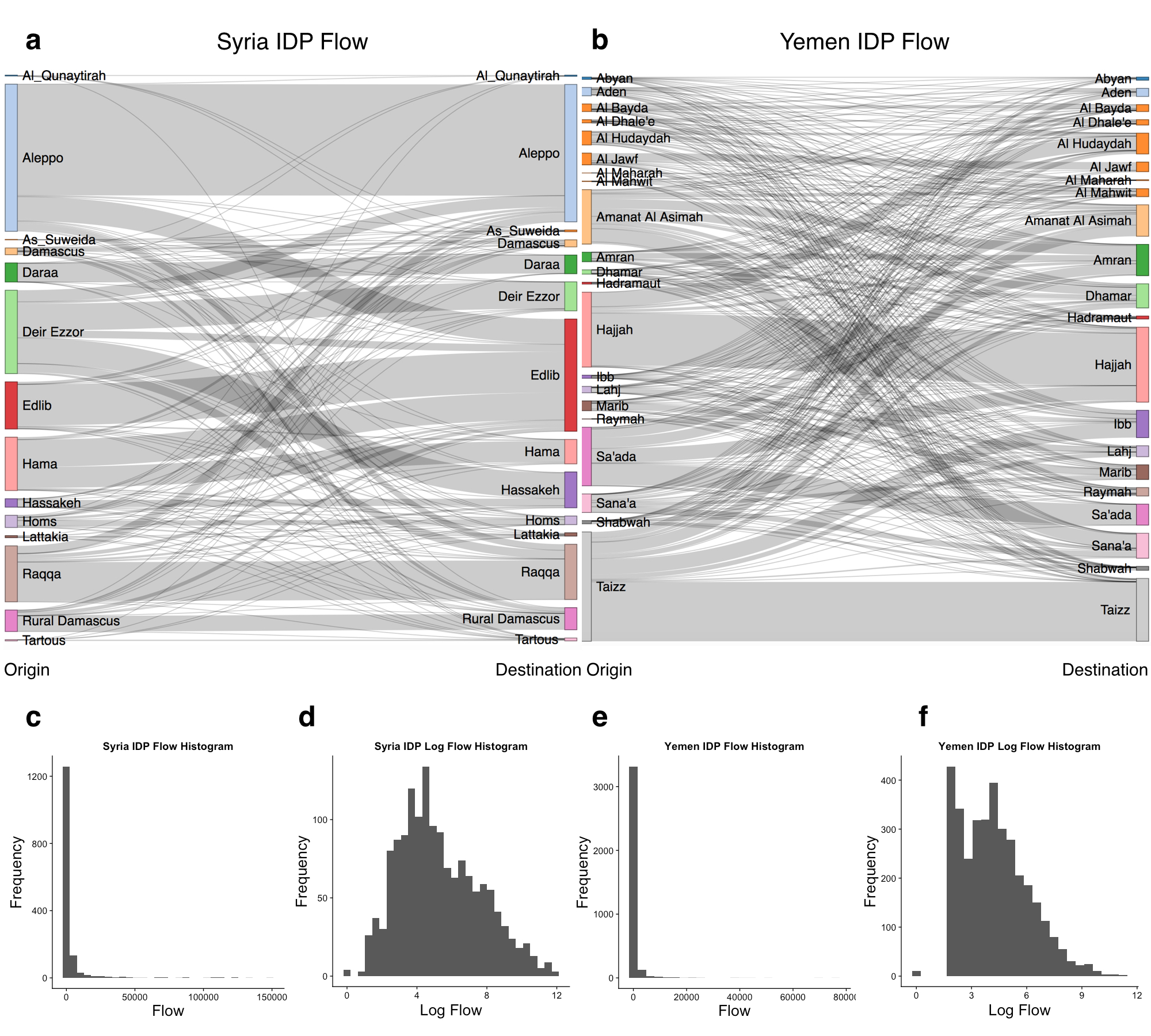}
\caption{a,b: IDP flow from province to province aggregated over all time periods for Syria (a) and Yemen (b). Each node represents a province. The widths of the bands represent the amount of flow. c-f: Distribution of IDP flows across all time points and provinces for Syria (c,d) and Yemen (e,f). Both log-transformed (d,f) and untransformed IDP flow values (c,e) are shown.}
\label{fig:sankeyLog}
\end{figure}

We observed that the distribution of IDP flow could be modeled as log-normal for the sake of forecasting large, rare displacement events (Extended Data Figure \ref{fig:sankeyLog}). Because our focus is on predictive performance and not statistical inference, the loss of effect size interpretability from log-transforming the response variable is not relevant. Furthermore, we empirically find that despite the bias introduced by transforming and untransforming the response variable, doing so provided better predictions than directly modeling IDP flow (Extended Data Table \ref{tab:logdiff}). Thus, all evaluation metrics for statistical models are reported from models trained on log flow, where predictions are untransformed back into flow. We ran our baseline persistence models both on log flow and normal flow separately, so as not to introduce bias from transforming and untransforming.
%
%
%
%
%
%
%

\begin{table}
\centering
\caption{Descriptive statistics on IDP arrivals, food prices, wages, fuel prices, and conflict intensity for Syria and Yemen. N denotes the number of observations and SD denotes standard deviation.  Wage data are unavailable for Yemen. Units for food/wage/fuel data are in Syrian and Yemeni currency, respectively.}
\resizebox{\columnwidth}{!}{%
\begin{tabular}{lrrrrrrrrrrr}
  \hline
Country & N & Flow Mean & Flow SD & Food Mean & Food SD & Wage Mean & Wage SD & Fuel Mean & Fuel SD & Conflict Mean & Conflict SD\\ 
  \hline
Syria & 1505 & 3098.59 & 12066.28 & 474.93 & 163.42 & 1383.18 & 373.97 & 2054.13 & 1077.89 & 0.36 & 1.33\\ 
Yemen & 3589 & 563.54 & 2912.12 & 280.04 & 155.13 &  &  & 977.83 & 1387.37 & 0.40 & 1.17\\ 
   \hline
\end{tabular}%
}
\label{tab:descriptive}
\end{table}

\begin{table}
\begingroup
\small
\caption{Predictive performance differences between models trained directly on IDP flow and models trained on log-flow (and then transformed back to flow). Positive values for RMSE and MAE and negative values for $R^{2}$ and sign accuracy indicate the model trained directly on flow performed worse by the given amount.}
\label{tab:logdiff}
\begin{subtable}{0.3\linewidth}
\centering
\caption{Syria}
\begin{tabular}{lrrrr}
  \hline
Model & RMSE & MAE & $R^{2}$ & Sign Acc \\ 
  \hline
LMM & 2136.71 & 573.64 & -0.18 & -0.13 \\ 
 SVM & -325.86 & 962.05 & 0.03 & -0.10 \\ 
 RF & 20.32 & 1119.93 & -0.10 & -0.11 \\ 
 MERF & 375.20 & 1112.60 & 0.01 & -0.11 \\ 
 XGB & 533.16 & 1310.00 & 0.03 & -0.09 \\ 
 MLP & 1555.76 & 1516.87 & -0.17 & -0.06 \\ 
 \hline
\end{tabular}
\end{subtable}%
\hspace{4cm}
\begin{subtable}{0.3\linewidth}
\centering
\caption{Yemen}
\begin{tabular}{rlrrrr}
  \hline
Model & RMSE & MAE & $R^{2}$ & Sign Acc. \\ 
  \hline
LMM & 127.53 & 12.08 & -0.10 & -0.16 \\ 
 SVM & 785.60 & 651.41 & -0.12 & -0.11 \\ 
 RF & -26.89 & 139.77 & 0.11 & -0.09 \\ 
 MERF & 272.47 & 202.70 & 0.06 & -0.11 \\ 
 XGB & 187.72 & 237.65 & -0.01 & -0.12 \\ 
 MLP & 1464.27 & 259.35 & 0.00 & -0.06 \\ 
   \hline
\end{tabular}
\end{subtable}
\endgroup
\end{table}

\bibliographystyle{plain}
\bibliography{sample}

\end{document}